\documentclass[preprint,12pt]{elsarticle}
\usepackage{etex}
\usepackage[export]{adjustbox}
\usepackage{tabularx}
\usepackage{placeins}
\usepackage{hyperref}
\usepackage{tikz}
\usepackage[pdftex]{lscape}
\usepackage{tikz-3dplot}
\usepackage{pgfplots}
\usepackage{tikz}
\usepackage{tikz-3dplot}
\usepackage{tikz}
\usepackage{tikz-3dplot}
\usepackage{pgfplots}
\usetikzlibrary{shapes,arrows}
\usepackage{setspace}
\usepackage{xspace}
\usepackage{pgfplotstable}
\usepackage{ifthen}
\usepackage{subfigure}
\usetikzlibrary{calc}
\usepackage{placeins}
\usepackage[titletoc,title]{appendix}
\usepackage{hyperref}
\usepackage{ifthen}
\usepackage{units}
\usepackage{rotating}
\usepackage{epsfig}
\usepackage[centertags,intlimits]{amsmath}
\usepackage{amssymb,amscd}
\usepackage{array}
\usepackage{multirow}
\usepackage{supertabular}
\usepackage{dcolumn}
\usepackage{xspace}
\usepackage{upgreek}
\usepackage{color}
\usepackage{calc}
\usepackage{ifthen}
\usepackage{caption}
\usepackage{cleveref}
\usepackage{color}
\usepackage[titletoc,title]{appendix}
\usetikzlibrary{intersections}
\usetikzlibrary{patterns}
\usepackage{lineno}
\usepackage{booktabs}
\usepackage{listings}
\usepackage[normalem]{ulem}

\definecolor{mygreen}{rgb}{0,0.6,0}
\definecolor{mygray}{rgb}{0.5,0.5,0.5}
\definecolor{mymauve}{rgb}{0.58,0,0.82}

\usepackage{listings}
\lstMakeShortInline|

\usepackage{natbib}

\newcommand*{\mybox}[1]{\uline{\texttt{\smaller #1}}\xspace}

\newcommand*{\myfile}[1]{\textsf{#1}\xspace}

\mathchardef\mhyphen="2D

\newcommand{\eVdist}{\kern-0.06667em}
\newcommand{\GeV}{\text{Ge\eVdist{}V\/}}
\newcommand{\TeV}{\text{Te\eVdist{}V\/}}
\newcommand{\MeV}{\text{Me\eVdist{}V\/}}
\newcommand{\CM}{\text{cm}}
\newcommand{\MM}{\text{mm}}

\usepackage{relsize}
\newcommand{\cxx}{\text{C\raisebox{0.1ex}{\smaller ++}}\xspace}
\newcommand{\PrintIDS}{
\begin{flushright}
MPP-2019-258, MCNET-19-27, LU-TP 19-58
\vspace{-0.2cm}
\end{flushright}
}

\journal{Computer Physics Communications}

\newcommand{\tabI}{
 \renewcommand{\arraystretch}{1.0}
\addtolength{\tabcolsep}{-4pt}
\begin{table} %[htbp]
\begin{center}
  \begin{tabular}{llccr}
    \toprule
    Operating system               &Repository                               & ROOT      & Version       & Credits    \\ %\hline\hline
    \midrule
Mac\,OS\,X                          &homebrew-hep~\cite{hephomebrew}         & no         & 3.2.0         &Enrico Bothmann  \\ %\hline
Arch Linux                            &AUR~\cite{aur}                           & no         & 3.1.1         &Frank Siegert    \\ %\hline
Debian\,9                        &Testing~\cite{debian}                    & no        & 3.1.2         &Mo Zhou          \\ %\hline
Ubuntu\,19                       &Universe~\cite{ubuntu}                   & no         & 3.1.1         &                 \\ %\hline
Fedora\,28+                      &EPEL~\cite{epel}                         & yes         & 3.2.0         &Mattias Ellert   \\ %\hline
RHEL\,7+ and like                &EPEL~\cite{epel}                         & yes         & 3.2.0         &Mattias Ellert   \\ %\hline
SUSE/openSUSE                  &Tumbleweed~\cite{opensusetumbleweed}     & no         & 3.1.1         &                 \\ %\hline
Linux                          &LCG~\cite{lcg}                           & yes         & 3.1.2         &                 \\ %\hline
Windows\,10                      &                                         & no         & 3.2.0         &                 \\ %\hline
BSD\,12                          &                                         & no         & 3.2.0         &                 \\ %\hline
Solaris                        &                                         & no         & 3.2.0         &                 \\ %\hline
Multiple                       &pypi~\cite{pypi}                         & no         & 3.2.0         & HepMC Devs.      \\ %\hline
Linux/MacOSX                   &conda-forge~\cite{conda}                 & no         & 3.2.0         & Henry Schreiner                \\ %\hline
    \bottomrule
  \end{tabular}
\end{center}
\caption{Summary on systems where HepMC3 was tested and the availability of HepMC3 precompiled binaries. For the majority of tests
only the Intel-compatible 64-bit architecture (x86\_64) was considered. The ROOT support was tested only for these
systems which provide ROOT packages in the repositories.  }
\label{tab:os}
\end{table}
}

\newcommand{\tabIIa}{
 \renewcommand{\arraystretch}{1.0}
\begin{table} %[htbp]
\begin{center}
  \begin{tabular}{llr@{\quad}r@{\quad}r}\toprule
  Code                                 & Type                     & \multicolumn{2}{c}{Matched versions} &  Interface    \\
  \cmidrule{3-4}
                                     &                          &  Code &   HepMC3    &  location   \\ %\hline\hline
  \midrule
SHERPA-MC~\cite{Bothmann:2019yzt}    & MCEG                     &$>$2.2.8 & 3.1+      &  SHERPA-MC      \\
                                     &                          &$>$2.2.6 & 3.0       &  SHERPA-MC  \\ \addlinespace %\hline
JetScape~\cite{Cao:2017zih}          & MCEG                     &   1.0   & 3.0       &       JetScape    \\ \addlinespace %\hline
ThePEG\,2~\cite{Lonnblad:2006pt}       & MCEG toolkit             & 2.2.0   & 3.1+      & ThePEG2\\ \addlinespace %\hline
Herwig\,7~\cite{Bellm:2015jjp}         & MCEG                     & 7.2.0   & 3.1+      & ThePEG2\\ \addlinespace %\hline
Pythia\,8~\cite{Sjostrand:2007gs}      & MCEG                     & 8.2+    & 3.X       &  HepMC3\\ \addlinespace %\hline
Pythia\,6~\cite{Sjostrand:2006za}      & MCEG                     & 6.4     & 3.1+      & HepMC3\\ \addlinespace %\hline
Tauola~\cite{Jadach:1990mz}          & MCEG                     & 1.1.6c  & 3.X       & HepMC3\\ \addlinespace %\hline
Photos~\cite{Barberio:1993qi}        & MCEG                     & 3.61    & 3.X       & HepMC3\\ \addlinespace %\hline
WHIZARD~\cite{Kilian:2007gr}         & MCEG                     & $>$2.8.1& 3.1+      &WHIZARD\\ \addlinespace %\hline
Rapgap~\cite{Jung:1993gf}            & MCEG                     & $>$3.303${}^{*}$     & 3.1+ &Rapgap \\ \addlinespace %\hline
Cascade~\cite{Jung:2010si}           & MCEG                     & $>$3.00${}^{*}$      & 3.1+ &Cascade \\ \addlinespace %\hline
EvtGen~\cite{Lange:2001uf}           & MCEG                     & master${}^{*}$       & 3.1+ &EvtGen\\ \addlinespace %\hline
Geant\,V~\cite{Amadio:2018tnh}         & Simulation               & master   & 3.0       &GeantV\\ \addlinespace %\hline
MC-TESTER~\cite{Davidson:2008ma}     & Testing                  & 1.25  & 3.X          &HepMC3\\ \addlinespace %\hline
  Rivet~\cite{Buckley:2010ar}          & Testing                  & 3.0.3 & 3.1+         &Rivet \\ \addlinespace %\hline
  \bottomrule
\end{tabular}
\end{center}
\caption{Summary on the usage of HepMC3 in external projects. ``master''
 stands for the latest version in the used  version control system of the official repository,
 e.g.\ for master branch of git repository. If known, the versions where
 support is expected to be released are given in brackets. The  ${}^{*}$  symbols denote support implemented in non-official versions of the codes.
}
\label{tab:summary}
\end{table}
}

\newcommand{\tabstatusp}{
 \renewcommand{\arraystretch}{1.0}
\begin{table}[htbp]
\begin{center}
  \begin{tabular}{l@{\quad}l@{\quad}l}
    \toprule
    Status code                              & Meaning                           & Usage    \\
    \midrule
    0  &  Not defined (null entry)           & Not a meaningful status \\ %\hline
    \addlinespace
    1  &  Undecayed physical particle        & Recommended for all cases \\ %\hline
    \addlinespace
    2  &  Decayed physical particle          & Recommended for all cases \\ %\hline
    \addlinespace
    3  &  Documentation line                 & Often used to indicate       \\
       &                                     &   in/out particles in hard process      \\ %\hline
    \addlinespace
    4  &  Incoming beam particle             & Recommended for all cases \\ %\hline
    \addlinespace
    5--10  & Reserved for future standards    & Should not be used  \\ %\hline
    \addlinespace
    11--200  & Generator-dependent            & For generator usage  \\ %\hline
    \addlinespace
    201--    & Simulation-dependent           & For simulation software usage  \\ %\hline
    \bottomrule
\end{tabular}
\end{center}
\caption{Status codes for particles.}
\label{tab:statusp}
\end{table}
}

\newcommand{\tabstatusv}{
 \renewcommand{\arraystretch}{1.0}
\begin{table}[htbp]
\begin{center}
\begin{tabular}{l@{\quad}l@{\quad}l}\toprule
  Status code                               &Meaning                           & Usage    \\
  \midrule
  0  &  Not defined (null entry)            & Vertex with no meaningful status \\ %\hline
  \addlinespace
  1-  &  Generator-dependent             & For generator usage  \\ %\hline
  \bottomrule
\end{tabular}
\end{center}
\caption{Status codes for vertices.}
\label{tab:statusv}
\end{table}
}

\newcommand{\tabIII}{
 \renewcommand{\arraystretch}{1.0}
\begin{table} %[htbp]
\begin{center}
\begin{tabular}{l@{\quad}l@{\quad}r}\toprule
Package or                                                  & Used in       & Purpose \\
  feature                                                                     &           &       \\  %\hline\hline
  \midrule
  ROOT\,6                                                               & ROOT, examples, tests          & Provide ROOT I/O            \\ \addlinespace
  Doxygen~\cite{doxygen}                                              & documentation          &   Generate documentation          \\ \addlinespace %\hline
  Pythia\,6                                                           & interfaces, examples           & Provide Pythia6 example             \\ \addlinespace %\hline
  Pythia\,8                                                            & interfaces, examples,    & Pythia8 examples and tests             \\
                                                              &  tests    &              \\ \addlinespace %\hline
  TAUOLA                                                             &  interfaces, examples,   & PHOTOS examples and tests             \\
                                                               &  tests   &              \\ \addlinespace %\hline
  PHOTOS                                                             &  interfaces, examples,    & Tauola examples and tests             \\
                                                              &  tests    &              \\ \addlinespace %\hline
  HepMC\,2                                                             &  tests         & Compare HepMC3 vs HepMC2            \\ \addlinespace %\hline
  threads                                                           &  tests         & Check thread safety            \\ \addlinespace %\hline
  graphviz~\cite{graphviz}                                          &  examples         & Provide GUI event viewer            \\ \addlinespace %\hline
  valgrind~\cite{valgrind}                                           &  tests         &  Check for memory leaks           \\ \addlinespace %\hline
  zlib~\cite{zlib}                                                   &  examples         &  Access compressed ASCII files           \\ \addlinespace %\hline
  Python~\cite{python27,python3}                                     &  Python, tests         & Compile/test Python bindings            \\ \addlinespace %\hline %\hline
  \addlinespace
  \midrule[0.1pt]
  \addlinespace
  \addlinespace
  binder~\cite{binder}                                               &  Python development         & Generate Python bindings            \\ \addlinespace %\hline
  astyle~\cite{astyle}                                               & development          & Format the code            \\ \addlinespace %\hline
  cppcheck~\cite{cppcheck}                                           & development          & Do static analysis of the code            \\ \addlinespace %\hline
  NSIS~\cite{nsis}                                            & development          &   Create Windows installers          \\ \addlinespace %\hline
  gengetopt~\cite{gengetopt}                                            & development          &   Create option parsers          \\ \addlinespace %\hline
  \bottomrule
\end{tabular}
\end{center}
\caption{Summary of the packages that can be used in HepMC3. The packages used for development only are given in the bottom part of the table.
}
\label{tab:deps}
\end{table}
}

\newcommand{\tabexamples}{
 \renewcommand{\arraystretch}{1.0}
\begin{table} %[htbp]
\begin{center}
\begin{tabular}{l@{\quad}l@{\quad}l}\toprule
  Example location                                                 & Requires       & Purpose \\ %\hline\hline
  \midrule
  \addlinespace
  \myfile{BasicExamples/}                                                               &          &           \\
  \hspace*{1em}\myfile{basic\_tree.cc}                                                               &          &  Build event from scratch            \\
  \hspace*{1em}\myfile{hepevt\_wrapper\_example\_fortran.f}                                             &FORTRAN   & Use HEPEVT wrapper            \\
  \hspace*{1em}\myfile{HepMC2\_reader\_example.cc}                                                     &          &  Read HepMC2             \\
                                                                                 &          &   IO\_GenEvent files            \\
  \hspace*{1em}\myfile{HepMC3\_fileIO\_example.cc}                                                     &          &  Read HepMC3 Asciiv3            \\ \addlinespace
  \addlinespace
  \myfile{ConvertExample/}                                                      & (ROOT,zlib)         &  Convert files from             \\
                                                                       &                     &  one format into another            \\ \addlinespace
  \myfile{LHEFExample/}                                                      &          &  Manipulate  LHEF events            \\ \addlinespace
  \myfile{Pythia6Example/}                                                      &  FORTRAN        &  Use Pythia6 interface            \\ \addlinespace
  \myfile{Pythia8Example/}                                                      &  Pythia8        &  Use Pythia8 interface            \\ \addlinespace
  \myfile{ViewerExample/}                                                      &  ROOT,                 & Use GUI event browser            \\
                                                                      &        graphviz        &              \\ \addlinespace
  \myfile{RootIOExample/}                                                      &  ROOT        &  Use ROOT format             \\ \addlinespace
  \myfile{RootIOExample2/}                                                      &  ROOT        &  Use  ROOT format            \\
                                                                       &              &    with own class          \\ \addlinespace
  \myfile{RootIOExample3/}                                                      &  ROOT        &  Use  ROOTTree format          \\ \addlinespace
  \bottomrule
\end{tabular}
\end{center}
\caption{List of examples in HepMC3. The optional dependencies are given in brackets.
}
\label{tab:examples}
\end{table}
}

\begin{document}
\begin{frontmatter}

\title{The HepMC3 Event Record Library\\ for Monte Carlo Event Generators}

\author[ab]{Andy~Buckley}
%\ead{andy.buckley@cern.ch}
\address[ab]{School of Physics \& Astronomy, University of Glasgow, Glasgow, UK}

\author[pi]{Philip~Ilten}
\address[pi]{School of Physics and Astronomy, University of Birmingham, Birmingham, UK}

\author[dk]{Dmitri~Konstantinov}
%\ead{dmitri.konstantinov@cern.ch}
\address[dk]{NRC Kurchatov Institute -- IHEP, Protvino, RU}

\author[ll]{Leif~L\"onnblad}
%\ead{leif.lonnblad@thep.lu.se}
\address[ll]{Department of Astronomy and Theoretical Physics, Lund University, Lund, SE}

\author[jm]{James~Monk}
\address[jm]{Formerly at Niels Bohr Institutut, Copenhagen, DK \emph{and} Lund University, Lund, SE}

\author[wp]{Witold Pokorski}
%\ead{witold.pokorski@cern.ch}
\address[wp]{CERN, Geneva, CH}

\author[tp]{Tomasz~Przedzinski}
%\ead{tomasz.przedzinski@cern.ch}
\address[tp]{Formerly at Jagiellonian University -- Institute of Physics, Cracow, PL}

\author[av]{Andrii Verbytskyi}
%\ead{andrii.verbytskyi@mpp.mpg.de}
\address[av]{Max-Planck-Institut f\"ur Physik, Munich, DE}

\vspace{-1em}

\begin{abstract}
In high-energy physics, Monte Carlo event generators (MCEGs) are used to
simulate the interactions of high energy particles. MCEG event records store the
information on the simulated particles and their relationships, and thus
reflects the simulated evolution of physics phenomena in each collision event.

We present the HepMC3 library, a next-generation framework for MCEG event record
encoding and manipulation, which builds on the functionality of its widely-used
predecessors to enable more sophisticated algorithms for event-record analysis.
By comparison to previous versions, the event record structure has been
simplified, while adding the possibility to encode arbitrary information. The
I/O functionality has been extended to support common input and output formats
of various HEP MCEGs, including formats used in Fortran MCEGs, the formats
established by the HepMC2 library, and binary formats such as ROOT; custom input
or output handlers may also be used. HepMC3 is already supported by popular
modern MCEGs and can replace the older HepMC versions in many others.

\end{abstract}


\end{frontmatter}
\noindent
{\bf PROGRAM SUMMARY}%Delete as appropriate.
\vspace{0.5cm}\\
\begin{small}
\noindent
{\em Manuscript Title:} The HepMC3 Event Record Library for Monte Carlo Event Generators \\
{\em Authors:} Andy Buckley, Philip Ilten, Dmitri Konstantinov, Leif L\"onnblad, James Monk, Witold Pokorski, Tomasz Przedzinski, Andrii Verbytskyi.\\
{\em Program Title:} HepMC\,3\\
{\em Licensing provisions:} GPLv3 \\
{\em Programming language:} C++  \\
{\em Operating system:} GNU/Linux, Mac OS X, Windows, Unix \\
%{\em Memory usage:} \\
%{\em External routines/libraries:} \\
{\em Nature of problem:} The simulation of elementary particle reactions at high energies requires to store and/or modify information related to the simulation.\\
%{\em Solution method:} \\
%{\em Running time:} \\
\end{small}

\pagenumbering{arabic}
\newpage
\flushbottom
\tableofcontents
\newpage

\section{Introduction}
During the simulation of elementary particle reactions at high energies by MCEGs
it is necessary to store and/or modify information related to the simulation, in
the form of calculation elements, intermediate particles, decay cascades, etc.
The main purpose of the HepMC3 event record library~\cite{hepmc3}
is to hold this information both on per-event and simulation-run bases, and to
facilitate manipulations upon it. In what follows, we first review the design
principles of HepMC3 and the challenges which motivated its development, then
turn to its technical implementation, and usage.

\section{Data and object model}

The logical structure of the information in the HepMC3 library follows the
typical convention of modern MCEGs, being
split into two parts: general information on the conditions during simulation
execution (which is typically common for a run of events), and the simulated
events themselves. The first part contains the description of used tools and
settings applied in MCEG and thus, partially prescribes the interpretation of
the simulated events. Here and below we call this data ``run information''.

Each event from the second part holds a link to the run information and itself
consists of ``particles'', ``vertices'' and additional information about the
event or constituent ``particles'' and ``vertices''. In this scheme the
``particles'' directly correspond to the physical particles and therefore
possess physical properties -- four momentum, flavour\footnote{The exact
  particle-flavour encoding is not enforced in the library code, for reasons of
  performance and flexibility as the standard continually evolves. However, the
  examples in this paper, and all known users of the HepMC library, follows the
  enumeration scheme described in PDG~\cite{PDG2018}.}, status\footnote{See
  App.~\ref{status} for details.}, etc.  The ``vertices'' do not have a specific
physical meaning and simply indicate the elementary transmutation of a set of
``incoming'' particles into a set of ``outgoing'' particles: this may be a
purely technical operation and hence should not overinterpreted. Typical
examples of such a transmutation are $1 \to 2$ radiative splittings, $2 \to 2$
scatterings, $1 \to 1$ momentum-recoil corrections, and $1 \to n$ decays.
Therefore, the vertices hold the lists of pointers to incoming and outgoing
particles, the position in space-time of the assumed interaction or decay (if
defined), and the status. The latter is an abbreviated physically meaningful
description of the transmutation, see App.~\ref{status} for details.

The described event record structure results in a certain relation between
particles and vertices. In a vertex, for each incoming particle the outgoing
particles are considered as ``children'' and for each outgoing particle the
incoming particles are considered as ``parents''. From these definitions the
wider terms ``ancestors'' and ``descendants'' are inferred by recursion,
e.g.~parents of parents of\dots.

The particles that act as graph edges between vertices typically have a
``production vertex'' where they came from, and an ``end vertex'', where they
undergo their next modification or interaction: the only exceptions to this rule
are the stable final-state particles which have no end vertex, with the (usually
two) incoming beam particles which
are assigned to a unique ``root vertex'' without incoming particles.

The HepMC3 event record can hold events with arbitrarily complex relations
between the particles and vertices. However, to avoid algorithmic problems, it
is expected the event structure will adhere to the following rules:
\begin{itemize}
\item All particles and vertices in the event should be connected with each
  other, e.g.\ the event should not contain dangling particles or vertices.
\item Cyclic relations where a particle can be its own ancestor should be
  avoided.
\item All vertices should have at least one outgoing particle.
\item All vertices but root vertex should have at least one incoming particle.
\item Vertices should have a meaningful or zero status code\footnote{See
    App.~\ref{status} for details.}:
  \begin{itemize}
  \item Particles with no end vertex should be assigned status $1$;
  \item The incoming particles should be assigned status $4$.
  \end{itemize}
\item The number of weights in the event should match the number of the names
  for weights in the run information.
\end{itemize}

The event's constituent particles and vertices are collectively referred to as
``objects''.  Inside the event these are enumerated with non-zero integer
numbers (objects IDs, or OID), while OID=0 is reserved for the event itself. For
the correctly composed event the OIDs should be deducible from the event
topology, i.e.\ the particles are sorted according to the event topology and
their indices correspond to their position in the sorted list\footnote{For the
  ordering to be unique, an ordering rule is needed for topologically identical
  particles such as e.g.\ the initial leptons in $e^{+}e^{-}\rightarrow
  hadrons$. Such a rule cannot cover all potential cases, but, using the
  particle types, their charge, invariant mass or other quantities it can cover
  \emph{practically} all physically meaningful cases.} and are positive. The indexes of vertices
correspond to the minimal index of their incoming particles taken with minus
sign\footnote{Therefore, the root vertex has no index and all its properties are
  stored in the event.} and are negative.

Any additional piece of information on the whole event, particles or vertices is
called an attribute and can be stored inside the event using character
representation and referred object OID.  There are some standard physical use
case for the attributes: information on the polarisation, color (for particles),
type of the interaction (for vertices), information on the used parton density
functions (PDFs), process cross-section etc. (for event).  As every object can
have multiple attributes, these are distinguished by their names, that should be
unique within corresponding object.  No restrictions are imposed on the number,
type or names of the attributes. However, the users should not use for their
custom attributes names reserved for the standard attributes.  For the events
the standard attributes are:
\begin{itemize}
\item |GenCrossSection| -- an attribute holding the information on the
  cross-sections on the processes in the event.  The description of this
  attribute is given in App.~\ref{attributes}.
\item |GenPdfInfo| -- an attribute holding the information on the used PDFs.
  The description of this attribute is given in App.~\ref{attributes}.
\item |GenHeavyIon| -- An attribute holding the information on the heavy ions in
  the incoming beams.  The description of this attribute is given in
  App.~\ref{attributes}.
\item |alphaQCD| -- an attribute holding the floating point value of QCD
  coupling constant.
\item |alphaQED| -- an attribute holding the floating point value of QED
  coupling constant.
\item |event_scale| -- an attribute holding the floating point value of event
  hard scale.
\item |mpi| -- an attribute holding the number of multiparticle interactions integer.
\item |signal_process_id| -- an attribute holding an integer number that
  characterises the signal process in the event.  As the exact numbering scheme
  is not not defined, the value is generator dependent, see
  Ref.~\cite{Sjostrand:2006za} as an example.
\item |signal_vertex_id| -- An attribute holding the index of the vertex signal
  process.
\item |random_states1|, |random_states2| \dots |random_statesN| -- arbitrary
  number of attributes holding the integer number states of random number
  generator in the beginning of event simulation. The numbering should start
  from one. No gaps in the numbering of these states are allowed.
\item |random_states| -- vector of integer numbers corresponding to the
  states of random number generator at the beginning of event simulation.
\item |cycles| -- an attribute holding an integer number to show the presence of
  cyclic relations in the event.  The events with tree-like structure should
  have this attribute equal to zero or don't have it at all.
\end{itemize}

\noindent
The attributes |alphaQCD|, |alphaQED|,  |random_states|, |signal_process_id|, |mpi|
 and |signal_vertex_id| typically present in the events that were
originally produced with the HepMC2 library.

\noindent
For the vertices the single standard attribute is:
\begin{itemize}
\item |weights| -- vector of floating point numbers which correspond to the
  weights assigned to this vertex.
\end{itemize}

\noindent
For the particles the standard attributes are:
\begin{itemize}
\item |flows| -- vector of integer numbers which correspond to the QCD color
  flow information. No encoding scheme of the colour flows is imposed by the
  library, but it is expected to comply with the rules in Ref.~\cite{PDG2018}.
\item |theta| -- an attribute holding the floating point value of the $\theta$
  angle for polarisation.
\item |phi| -- an attribute holding the floating point value of the $\phi$ angle
  for polarisation.
\end{itemize}
If these attributes are present in the event they will be handled where it is
required, e.g.~in the event serialisation or in the interfaces to generators.
The implementation of the attributes is slightly different between the HepMC3
version 3.2.0 and the versions 3.1.x.  See section Sec.~\ref{compatibility} for
details.

\section{Implementation}
\label{impl}
Thanks to the usage of features of recent \cxx standards~\cite{cpp11}, the \cxx
implementation of the library has been significantly simplified with respect to
HepMC2. Many custom types and iterators were removed and the library became more
modular, allowing the implementation of custom features without breaking the
compatibility with core library components.

For efficient memory management most of the basic types are now used via the
smart pointers~\cite{smartpointers} as implemented in the \cxx standard
library. In addition, the concept of const-correctness~\cite{constcorrectness}
is promoted in the implementation of the library, fixing longstanding problems
where traversing the particle--vertex links in the event graph would permit a
|const| event event to be modified without resorting to use of
|const_cast|. Other defects, such as needing to obtain a non-const version of an
event in order to perform certain read-only operations have also been fixed in
HepMC3. To preserve this consistency, |const| versions of the HepMC3 smart
pointers are also implemented.

The main constituent classes of the library are briefly described below.

\subsection{C++ storage classes}
%\subsection{\cxx storage classes} Special commands in titles are complicated.
In  HepMC3 the information is represented via \cxx objects and
can be serialised as \cxx structures with plain data types.
The main types of objects (plain structures) in HepMC3 are:
\begin{itemize}
\item \lstinline{FourMomentum} -- a type that implements four vector in
  Minkovski space. The class includes some static functions for calculations of
  distance between vectors, their scalar product and other related quantities.
\item \lstinline{GenRunInfo } -- the main bookkeeping type that holds
  meta-information about the generated events: list of used tools, names of used
  event weights and arbitrary attributes. The embedded structure
  \lstinline{struct GenRunInfo::ToolInfo} (three \lstinline{std::string} fields)
  holds name, version and description of tool used for event generation and/or
  processing. This object can be serialised into plain data type structure
  \lstinline{GenRunInfoData}. The corresponding smart pointer types are
  \lstinline{GenRunInfoPtr} and \lstinline{ConstGenRunInfoPtr}.
\item \lstinline{GenEvent} -- the data type that holds the position of the
  primary interaction, and lists of vertices, particles and attributes. This
  object can be serialised into the plain data type structure
  \lstinline{GenEventData}. The relations between the particles and vertices are
  implemented in the \lstinline{GenEventData} structure as two lists of object
  OIDs. The relations between vertices and particles in \lstinline{GenEventData}
  are encoded via members \lstinline{std::vector<int> links1} and
  \lstinline{std::vector<int>links2} in a graph-like structure. The positive
  elements in \lstinline{std::vector<int> links1} stand for particles and that
  have end vertex OID encoded at the same position in
  \lstinline{std::vector<int> links2}. The negative elements in
  \lstinline{std::vector<int> links1} stand for production vertex with outgoing
  particle OID encoded in the same position in \lstinline{std::vector<int> links2}.
\item \lstinline{GenVertex} -- type of the objects used to describe
decays and interactions, holds its position, list of incoming and
outgoing particles, can have multiple attributes stored in the parent
\lstinline{GenEvent}. This object can be serialised into plain data
type structure \lstinline{GenVertexData}. The corresponding smart
pointer types are \lstinline{GenVertexPtr} and
\lstinline{ConstGenVertexPtr}.
\item \lstinline{GenParticle} -- type of objects used to describe
 particles, holds momenta, flavour, status of the particle, can have
multiple attributes stored in the parent |GenEvent|. This object
can be serialised into plain data type structure
\lstinline{GenParticleData}. The corresponding smart pointer types are
\lstinline{GenParticlePtr} and \lstinline{ConstGenParticlePtr}.
\item \lstinline{Attribute}  -- base class used to store arbitrary
information. The attribute data is stored as (and can be serialised to)
\lstinline{std::string}, which is used to initialise an object of
arbitrary type derived from the \lstinline{Attribute} class.
\end{itemize}

The \lstinline{Attribute} objects allow custom information to be stored in
the events.  Apart from the attributes used to store plain types (\lstinline{double, int, std::string}) and  the corresponding  vectors (\lstinline{std::vector<double>, std::vector<int>, std::vector<std::string>}) the
library provides implementation for the  \lstinline{GenPDFInfo}, \lstinline{GenCrossSection} and \lstinline{GenHeavyIon} attributes.
These are described in detail in App.~\ref{attributes}.

\subsection{Manipulation with objects}
The set of orthogonal operations is built in a way that objects manipulates on their constituents/subordinates and not vice verse.
The following basic operations are present in the HepMC3
\begin{itemize}
\item adding/removing particle to/from event.  The particle is added to the list of particles in the event if it is not present there already.
 While removing the particle attributes are removed as well. It is not checked if particle already belongs to any other event.\\
These functions are implemented  in\\
\lstinline{void GenEvent::add_particle(GenParticlePtr)} and in \\ \lstinline{void GenEvent::remove_particle(GenParticlePtr)}.
\item adding/removing particle to/from vertex.  The particle is added to the the list of vertex incoming or outgoing particles.
The production/end vertex of the particle is updated. In case the vertex belongs to an event, the particle will be added to the event as well.\\
These functions are implemented  in\\
\lstinline{void GenVertex::add_particle_in (GenParticlePtr)},\\ \lstinline{void GenVertex::add_particle_out(GenParticlePtr)},\\
\lstinline{void GenVertex::remove_particle_in (GenParticlePtr)} and\\ \lstinline{void GenVertex::remove_particle_out(GenParticlePtr)}.
\item adding/removing vertex to/from event. The vertex  and all it's particles are added to the list of event vertices/particles.
These functions are implemented  in \lstinline{void GenEvent::add_vertex(GenVertexPtr) } and in\\ \lstinline{void GenEvent::remove_vertex(GenVertexPtr)}.
\item adding/removing object attributes.  \\
These functions are implemented  in\\
\lstinline{bool GenEvent::add_attribute(const std::string&, std::shared_ptr<Attribute>)},\\
\lstinline{bool GenVertex::add_attribute(const std::string&, std::shared_ptr<Attribute>)},\\
\lstinline{bool GenParticle::add_attribute(const std::string&, std::shared_ptr<Attribute>)},\\
\lstinline{void GenEvent::remove_attribute(const std::string&)},\\
\lstinline{void GenParticle::remove_attribute(const std::string&)} and\\
\lstinline{void GenVertex::remove_attribute(const std::string&)}.
\item setting/getting the properties of run info, event, particles, vertices. For the full list of these functions we refer to the reference manual which is shipped with
the library and to the online reference manual~\cite{PDG2018}.
\end{itemize}
For a more convenient usage multiple basic functions were combined to operate on list of particles or vertices are implemented.
\subsection{LHEF classes}
\label{lhef}

Another important innovation in the HepMC3 library
is built-in support of routines for the LHEF event record/file
format~\cite{Boos:2001cv,Andersen:2014efa}.  The Les Houches
Event File format (LHEF) is used for passing events from a matrix-element
generator program (MEG) to a MCEG implementing parton showers, underlying event
models, hadronisation models etc.  Previously the standard implementation in
\cxx of the LHEF routines had already been maintained by Leif
L\"{o}nnblad. After the merger of the standard LHEF implementation into the
HepMC3 library, HepMC3 is a single package for manipulations with event records
used in MCEGs and MEGs.

\subsection{I/O classes and formats}
The serialisation of the MCEG event record is the most important part of the
library. Historically the serialisation was implemented in different packages
and in different formats.  The number of formats led to compatibility problems
in the interaction between different simulation packages.  For instance,
significant technical difficulties arise when the LHC-era MCEGs are used in the
simulation and reconstruction chains of older experiments~\cite{Akopov:2012bm}.  To overcome such
difficulties the reading and writing of events from/to disk was implemented in
classes that inherit from the same abstract classes
\lstinline{HepMC3::Reader/HepMC3::Writer}. Both base classes have very similar
structure. Apart from constructors and destructors only the following functions
are expected to be re-implemented:
\begin{itemize}
\item The method to fill next event from input\\   \lstinline{bool Reader::read_event(GenEvent& evt)}
\item The method to write event\\ \lstinline{void Writer::write_event(const GenEvent &evt)}
\item The methods to get input/output source state\\ \lstinline{bool Reader::failed()}/\lstinline{bool Writer::failed()}
\item The methods to close input/output source\\ \lstinline{bool Reader::close()}/\lstinline{bool Writer::close()}
\item The method to skip full reading some number of events\\ \lstinline{bool Reader::skip(const int n)}
\item The methods to  set/get  extra options for the I/O classes
\lstinline{void Reader::set_options(std::map<std::string, std::string>&)}, \lstinline{std::map<std::string, std::string> Reader::get_options() const},
\lstinline{void Writer::set_options(std::map<std::string, std::string>&)} and \lstinline{std::map<std::string, std::string> Writer::get_options() const}.
\end{itemize}

The standard methods to access \lstinline{GenRunInfo} objects that are used for
readers/writers are: \lstinline{std::shared_ptr<GenRunInfo> run_info()} and
\lstinline{void set_run_info(std::shared_ptr<GenRunInfo> run)}.
With such a design the algorithms to read or write events from/to external sources
are universal for all event formats, e.g.~for reading,\\[1ex]
\scalebox{.90}{
\begin{lstlisting}[label=lst:1,%caption=Reading of events in HepMC3 custom reader,
  language=C++,frame=lines,breaklines=false]
#include "MyCustomReader.h"
...
std::shared_ptr<Reader> examplereader;
examplereader= std::make_shared<MyCustomReader>(/*...*/);
...
while ( !examplereader->failed() ) {
  GenEvent evt(Units::GEV,Units::MM);
  examplereader->read_event(evt);
  if ( examplereader->failed() ) {
    std::cout << "End of file reached. Exit." << std::endl;
    break;
   }
}
\end{lstlisting}}\\[1ex]
In addition to the supported standard described formats,
the library allows users to implement customised input or output format via
implementation of custom \lstinline{Reader} and/or \lstinline{Writer} classes
inherited from the base classes \lstinline{Reader} and \lstinline{Writer}.  The
custom \lstinline{Reader} or \lstinline{Writer} class can be linked to the user
codes directly, either as in the previous code listing,
or used at run-time via a plugin mechanism:\\[1ex]\\
\scalebox{.90}{
\begin{lstlisting}[label=lst:2,%
  %caption=Reading of events in HepMC3 using custom plugin,
  language=C++,frame=lines,breaklines=false]
...
std::shared_ptr<Reader> examplereader;
examplereader= std::make_shared<ReaderPlugin>(input,
                            "libMyReader", "newMyReader");
...
while ( !examplereader->failed() ) {
  GenEvent evt(Units::GEV, Units::MM);
  examplereader->read_event(evt);
  if ( examplereader->failed() ) {
    std::cout << "End of file reached. Exit."<< std::endl;
    break;
  }
}
\end{lstlisting}}\\[1ex]

The supported formats described were introduced by different groups of people,
and for different purposes. Therefore the amount of information they hold is
significantly different.  The |ROOTTree|, |ROOT|, |LHEF| and |Asciiv3| formats,
in addition to the standard content, can hold almost arbitrary information via
the attributes mechanism.

\subsubsection*{IO\_GenEvent}
|IO_GenEvent| is an outdated text-based format used in the
HepMC2~\cite{Dobbs:2001ck} library.  The HepMC3 implementation is fully
compatible with that in the HepMC2 library.  However, unlike HepMC2, the reading
ends after the first occurring footer
|HepMC::IO_GenEvent-END_EVENT_LISTING|.

The |IO_GenEvent| record has fixed format, i.e.\ the information is limited to
particles, vertices, weights, PDF and heavy-ion information, and no extension is
allowed.

The attributes were used to reach compatibility with the HepMC2 software in the
I/O \lstinline{ReaderAsciiHepMC2} and \lstinline{WriterAsciiHepMC2} classes,
e.g.\ the attributes with names |alphaQCD| and |alphaEM| emulate the
corresponding class members of \lstinline{GenEvent} class in the HepMC2 library.
With this emulation the events can be read from |IO_GenEvent| files produced by
the HepMC2 library without any loss of information.

The classes that implement I/O in this format are \lstinline{ReaderAsciiHepMC2}
and \lstinline{WriterAsciiHepMC2}.  The reading of the events by the
\lstinline{ReaderAsciiHepMC2} can be tuned by the options
\lstinline{"vertex_weights_are_separated"},\\
\lstinline{"event_random_states_are_separated"} and
\lstinline{"particle_flows_are_separated"} -- see Sec.~\ref{compatibility} for
details.

\subsubsection*{Asciiv3}
|Asciiv3| is the HepMC3 native plain text format.  While being similar to
|IO_GenEvent|, this format is extendable and in comparison to the former requires
less storage space, as it does not save meaningless information on particles
(e.g.\ colour flow for hadrons).

The information on events is given between the header lines\\
|HepMC::Version X.Y.Z|\\
|HepMC::Ascii3-START_EVENT_LISTING|,\\
where X.Y.Z stands for library version
and the footer line\\ |HepMC::Ascii3-END_EVENT_LISTING|.\\
The run information (\lstinline{GenRunInfo}) is written after the header lines
followed by the lines with information on events.  Each non-empty line should
start from a one letter tag that defines how the content of the line should be
interpreted.  While reading\footnote{In the presented implementation the event
  might be omitted with \lstinline{bool Reader::skip(const int)} function
  without checks for correctness of tags. }
all unknown tags are treated as errors. The tags for the run information are ``W'', ``N'' and ``T''. These are used as follows:\\[1ex]
\mybox{W} \mybox{number of weights}\\
\mybox{N} \mybox{name of weight 1} \mybox{name of weight 2} \dots\\
\mybox{T} \mybox{name of tool 1}  \mybox{version of tool 1} \mybox{description of tool 1}\\[1ex]
The tag ``T'' can appear multiple times.\\
Each event starts from line with leading character ``E'' and ends with the next
line with leading character ``E'' or footer line.
The following tags are parsed:\\[1ex]
  \mybox{E} \mybox{number of particles} \mybox{number of vertices}\\
  \mybox{W} \mybox{value of weight 1} \mybox{value of weight 2} \dots\\
  \mybox{U} \mybox{momentum unit} \mybox{length unit}\\
  \mybox{A} \mybox{object OID} \mybox{attribute name} \mybox{string 1} \mybox{string 2} \mybox{string 3} \dots\\
  \mybox{P} \mybox{particle OID} \mybox{parent vertex OID} \mybox{PDG I.D.}  \mybox{$p_{x}$} \mybox{$p_{y}$} \mybox{$p_{z}$} \mybox{$e$} \mybox{particle mass}    \mybox{status},\\[1ex]
where $p_{x}$, $p_{y}$, $p_{z}$ and ${e}$ stand for the particle $4$-momentum
components.
If the production vertex has only one incoming particle, the outgoing particles can be presented as  \\[1ex]
  \mybox{P} \mybox{particle OID} \mybox{parent particle OID} \mybox{PDG I.D.}  \mybox{$p_{x}$} \mybox{$p_{y}$} \mybox{$p_{z}$} \mybox{$e$} \mybox{particle mass}    \mybox{status}\\
  \mybox{V}  \mybox{vertex OID}  \mybox{status}  \mybox{(comma-separated list of incoming OIDs)}   \mybox{@}   \mybox{$x$}  \mybox{$y$}  \mybox{$z$}  \mybox{$t$},\\[1ex]
  where ${x}$, ${y}$, ${z}$ and $t$ stand for the correspond position components of
the vertex and production time.
In case all components of the vertex position are zero, these can be omitted  \\[1ex]
  \mybox{V}  \mybox{vertex OID}  \mybox{status}  \mybox{(comma-separated list of incoming OIDs)}.\\[1ex]
  The tags ``E'', ``W'', ``U'' should appear only once per event.  Multiple
``A'', ``P'', ``V'', ``T'' tags per event are allowed.  Note that vertex with no
position and zero status will not appear in the listing explicitly.

The classes that implement I/O in this format are \lstinline{ReaderAscii} and
\lstinline{WriterAscii}.

\subsubsection*{HEPEVT} HEPEVT is an outdated plain text based format used by
many MCEGs written in Fortran (e.g.\ Pythia6).  The main purpose of the
implementation is to provide a compatibility layer for the MCEGs used in the
completed HEP experiments at HERA, LEP and PETRA machines.  The HEPEVT is the
most restrictive format and holds only the information on the particles without
any options for extra information. A more detailed description can be found
elsewhere~\cite{Altarelli:1989hx}.  The classes that implement I/O in this
format are \lstinline{ReaderHEPEVT} and \lstinline{WriterHEPEVT}.  The reading
of the events by the \lstinline{ReaderHEPEVT} can be tuned with an option
\lstinline{"vertices_positions_are_absent"}. The option should be present in the
list of options of the \lstinline{ReaderHEPEVT} object to read event record
without vertex positions.

\subsubsection*{ROOTTree} ROOTTree is a binary format based on the
ROOT~\cite{Antcheva:2009zz} \lstinline{TTree}.  This format is implemented using
customisation of ROOT \lstinline{Streamer} class.  Basically, objects of
interests (e.g.\ \lstinline{GenEvent}, \lstinline{GenParticle} and others) are
serialised as into corresponding data structures (e.g.\
\lstinline{GenEventData}, \lstinline{GenParticleData}) and written in this way
as branches of ROOT \lstinline{TTree}. As a result, the corresponding
\lstinline{TTree} saved to a ROOT file, can be interpreted with standard ROOT
without the HepMC3 library itself, i.e.\ a user with standard ROOT can retrieve
all information on the events in a form of simple structures
\lstinline{GenEventData}, \lstinline{GenParticleData} etc.

This has several advantages in comparison to the other formats: it allows random
access, access over network, has the best I/O performance and requires the
smallest amount of storage space per event.  The classes that implement I/O in
this format are \lstinline{ReaderROOTTree} and \lstinline{WriterROOTTree}.

\subsubsection*{ROOT} ROOT is a binary format based on the
ROOT~\cite{Antcheva:2009zz}.  This format is implemented using standard ROOT
serialisation and writes the objects to ROOT files ``as is''.  The classes that
implement I/O in this format are \lstinline{ReaderROOT} and
\lstinline{WriterROOT}.

\subsubsection*{LHEF}
The plain-text Les Houches Event Format, primarily intended for low-multiplicity
partonic matrix-element event communication.  The class that implement I/O in
this format is \lstinline{ReaderLHEF}. Currently no implementation of \lstinline{Writer}
is provided.  The documentation on the LHEF functions can be found
elsewhere~\cite{Andersen:2014efa}.

\subsection{Search module classes}
HepMC3 comes with an optional ``search'' library for finding particles
related to other particles or vertices.  Two main interfaces are
defined: Relatives, for finding a particular type of relative, and
Feature, for generating filters based on Features extracted from
particles.  In addition, the standard boolean operator on Filters are
also defined.  A \lstinline{Filter} is any object that has an operator
that takes as input a \lstinline{ConstGenParticlePtr} and returns a
\lstinline{bool} that reflects whether the input particle passes the
filter requirements or not.  Filter is defined in \myfile{Filter.h} as
an typedef of \lstinline{std::function<bool(ConstGenParticlePtr)>}.
The filters may use the \lstinline{Selector} class to extract standard
features from a particle and construct relational filters. As an
illustrative example the following code will obtain a list of all
final state descendants of a particle that has a transverse momentum
larger than 0.1~\GeV{} and has a pseudorapidity between -2.5 and 2.5:\\[1ex]
\scalebox{.90}{
\begin{lstlisting}[label=lst:1,
  % caption=Example use of the optional search library in HepMC3,
  language=C++,frame=lines,breaklines=false]
std::vector<ConstGenParticlePtr>
getDescendants(ConstGenParticlePtr parent) {
  Filter f = (StandardSelector::STATUS == 1 &&
              StandardSelector::PT > 0.1 &&
              StandardSelector::ETA > -2.5 &&
              StandardSelector::ETA < 2.5);
  return applyFilter(f, Relatives::DESCENDANTS(parent));
}
\end{lstlisting}}\\[1ex]

\subsection{Other classes and free functions}
In addition to the classes described above, HepMC3 includes a small number of
auxiliary classes.

The \lstinline{Setup} class controls verbosity of warnings.

The \lstinline{Units} class holds information on used units. The allowed length
units are \MM{} and \CM{}, while the allowed energy units are \MeV{} and~\GeV{}. The function %
\lstinline{GenEvent::set_units(Units::MomentumUnit, Units::LengthUnit)} %
performs conversion between different units used in the
event. Note that it does not affect the units used in the attributes of event.

The \lstinline{Print} class provides multiple static functions to produce
human-readable printings of objects in the library.  The same task is performed
with free overloaded operators \lstinline{<<} in \myfile{PrintStreams.h} header.

The functions and macros that help to find out the version of library are
located in \myfile{Version.h} header.

The header \myfile{ReaderFactory.h} provides  functions
\lstinline{std::shared_ptr<Reader> deduce_reader(const std::string &filename)} and
\lstinline{std::shared_ptr<Reader> deduce_reader(std::istream &)}
that try to open the a file or stream for reading and automatically deduce the
appropriate reader.

\section{Installation, dependencies, compatibility  and usage}
HepMC3 supports GNU/Linux, OS~X and Windows operation systems and
should be able to operate on some other Unix systems.
  It has been
tested on Ubuntu, CentOS, Fedora, openSUSE, Windows\,10 and OS~X operating
systems on Intel-compatible 64-bit processors.  Binary packages are available
for multiple operating systems, see Tab.~\ref{tab:os} for details.

%TODO: float table
\tabI

HepMC3 may be installed either from source, or by using precompiled packages
from the repositories of corresponding Linux distributions (for Linux users), or
from Homebrew-HEP for OS~X users. For the Windows, BSD and Solaris users it is
necessary to build the library from sources. Windows\,10 users should be able
create NSIS~\cite{nsis} installers if needed.
Python-based users can install the HepMC3 packages from the
CondaForge~\cite{conda} or PyPI~\cite{pypi} repositories.

The detailed instructions to compile the library from sources are provided in
the \myfile{README.md} file distributed with the library source codes and are
the same for all the supported platforms.  Only a short version is given below.

\subsection{Dependencies}
The only basic dependency for the installation of the library from sources is
the availability of a \cxx11 compatible \cxx compiler with appropriate run-time
and the build tool CMake~\cite{cmake}. It is recommended to use CMake of version
3.9 and newer.  The basic features of the package can be extended if additional
packages are available, see Tab.~\ref{tab:deps}.

%TODO: float table
\tabIII

\subsection{Installation from sources}
The procedure of installation from sources consists of multiple
steps\footnote{Here and below the commands are given assuming POSIX-compatible
  shell (e.g.\ GNU bash) and Unix-like OS.}.  The first step is to get the
HepMC3 sources from the git~\cite{git} repository:
\begin{lstlisting}[language=bash]

git clone https://gitlab.cern.ch/hepmc/HepMC3.git
\end{lstlisting}
or from the official site:
\begin{lstlisting}[language=bash]

wget http://cern.ch/hepmc/releases/HepMC3-3.2.0.tar.gz
tar -xzf HepMC3-3.2.0.tar.gz
\end{lstlisting}
Windows users can use web-browsers and/or proprietary utilities instead.
The second step is to create a work-space area on which to perform the builds:
\begin{lstlisting}[language=bash]

mkdir myhepmc3-build
cd myhepmc3-build
\end{lstlisting}
The third step is to configure, build and install the code with
CMake~\cite{cmake}\footnote{CMake of version 3 could be named as ``cmake3'' on
  some systems.}, e.g.\
\begin{lstlisting}[language=bash]

cmake -DCMAKE_INSTALL_PREFIX=../MyInstallationLocation  -DHEPMC3_ENABLE_ROOTIO=OFF  ../HepMC3
cmake --build ./
cmake --install ./
\end{lstlisting}
Optionally, after the compilation, it is possible to run the build-in test suite
based on CTest~\cite{cmake}:
\begin{lstlisting}[language=bash]

ctest  ./
\end{lstlisting}

\subsection{Compatibility}
Starting from version 3.1.0, the HepMC3 and HepMC2 libraries can co-exist in one
installation, therefore the migration of user code from HepMC2 to HepMC3 can go
as easy as possible.
\subsection{Usage}
As of end 2019 several MCEGs were interfaced to HepMC3, see
Tab.~\ref{tab:summary} for details.

% TODO: float table
\tabIIa

\FloatBarrier

\section{External codes}
\noindent The library itself embeds some external codes. These are:
\begin{itemize}
\item pybind11~\cite{pybind11}, a header-only library  used for python bindings.
\item Pythia\,6~\cite{Sjostrand:2006za}, a MCEG generator used in the examples.
\item gzstream~\cite{gzstream}, a set of \cxx  classes wrapping the zlib compression library.
\item Codes from examples of the binder~\cite{binder} package.
\item Various cmake modules were taken from the cmake distribution, see details in the corresponding modules.
\end{itemize}
The initial version of the Pythia\,8 HepMC3 interface was committed by Mikhail Kirsanov, who created the HepMC2
interface for the Pythia\,8 package~\cite{Sjostrand:2007gs}. The later versions were improved by Philip Ilten.

\section{Performance}
During the event generation by the MCEGs the speed of event construction
typically is not of great concern.  Moreover, it strongly depends on the type of
generator, its settings and therefore is not well defined.  Therefore, we
concentrate on a better defined characteristics of I/O performance while using
already generated events.  The input samples~\cite{testsamples} consist of
multiple event samples with various signal processes saved in HepMC2 files.
These include the $e^+e^- \to \text{hadrons}$ processes for
$\sqrt{s}=10$--$206~\GeV$, $e^+e^-\to \Upsilon$, $e^{\pm}p$ deep-inelastic
scattering, $pp \to \text{jets}$ for $\sqrt{s}=7~\text{and}~13~\TeV$, and more.

With these samples series of tests were performed with HepMC2 and HepMC3
libraries. All tests were performed on CentOS\,7 x86\_64 with ROOT version 6.18,
zlib version 1.27, HepMC2 version 2.06.10, gcc version 4.8.5 and default
settings for ROOT compression level, ROOT compression algorithm and the
precision of |Asciiv3| output.  Before the tests all the files were loaded into
memory.

\begin{figure}[htbp]\centering
\includegraphics[width=1.0\textwidth]{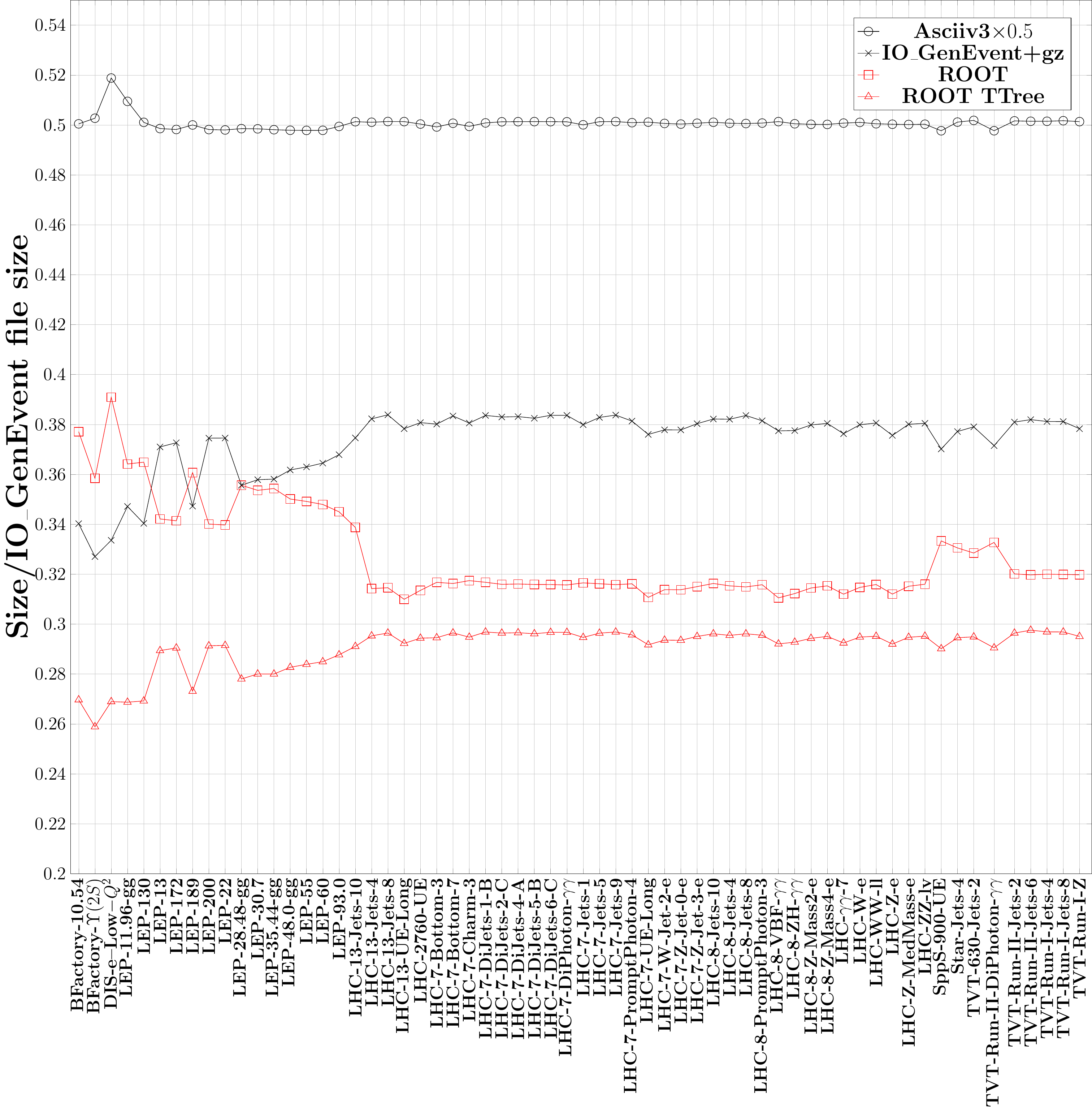}
\caption{Size of events samples in different formats relatively to the size of
  same sample in HepMC2 \texttt{IO\_GenEvent} format.  The physical content of the files
  with simulated events is encoded in the name of file.  ``BFactory'' and
  ``LEP'' in the file names indicate simulation of $e^{+}e^{-}$ collisions at
  $B$-factories and PETRA/TRISTAN/LEP colliders.  The main simulated processes
  are $e^{+}e^{-} \to \text{hadrons}$ for ``LEP'' and
  $e^{+}e^{-}\rightarrow \text{resonances} \rightarrow \text{hadrons}$.  ``DIS'' in the file
  name indicates the simulation of deeply-inelastic $e^{\pm}p$ scattering at
  HERA collider.  ``LHC'', ``SppS'' and ``TVT'' in the file name indicate the
  simulation of $pp$ collisions at LHC, SppS or Tevatron colliders.  The numbers
  following the collider name abbreviate the centre-of mass energy of the
  collision in \GeV{} or \TeV{}.  In addition, the names of files with $pp$
  simulated events include the abbreviated in the main process name, e.g.\
  ``LHC-8-Jets'' abbreviates the inclusive jet production.  }
\label{fig:size}
\end{figure}

The measurements of relative samples sizes are given in Fig.~\ref{fig:size}.

\begin{figure}[htbp]\centering
\includegraphics[width=1.0\textwidth]{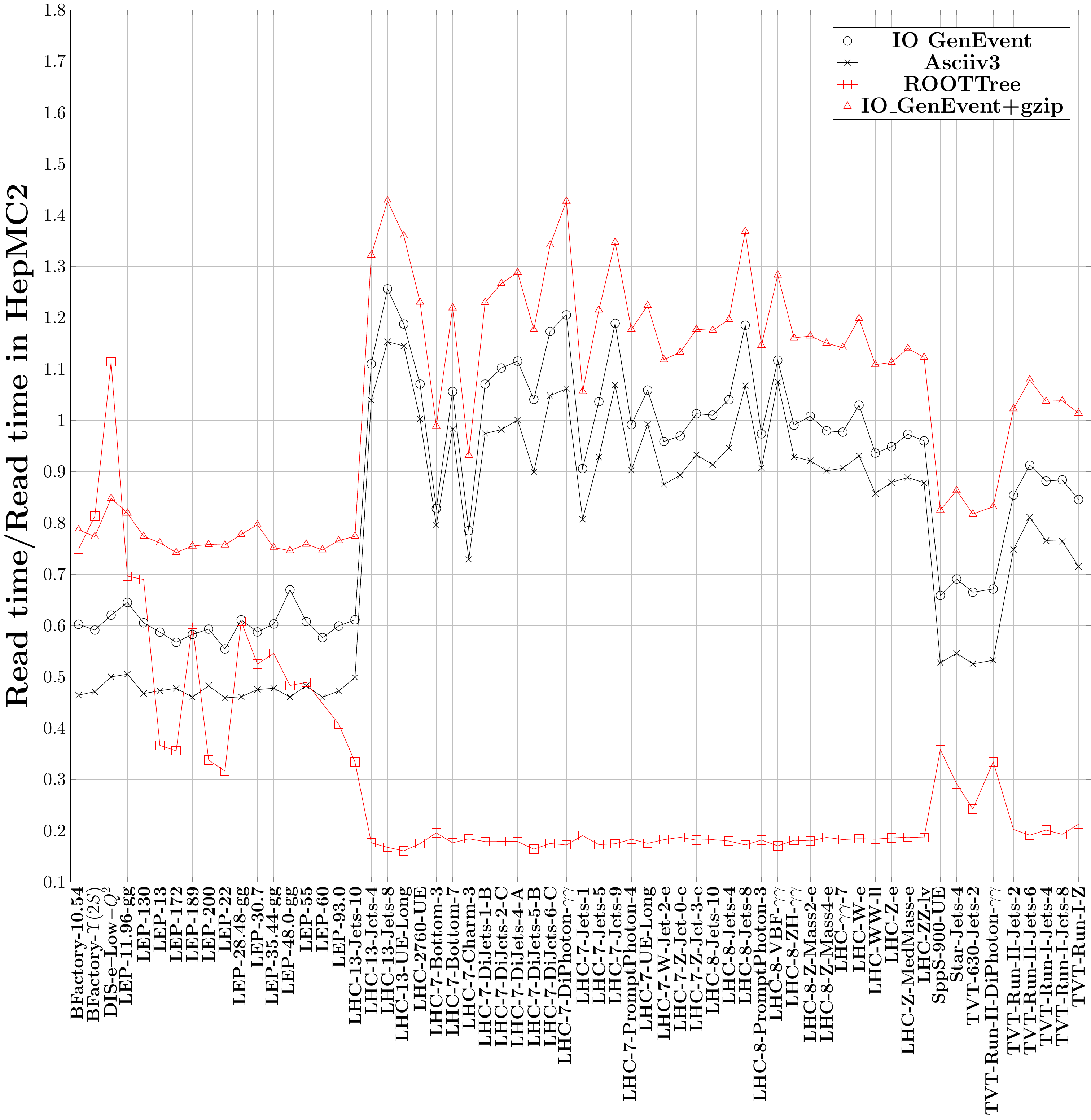}
\caption{Total reading time  of events samples in different formats relatively to the total reading time of same sample in HepMC2 IO\_GenEvent format.  See Fig.~\ref{fig:size} for details.}
\label{fig:totaltime}
\end{figure}

\begin{figure}[htbp]\centering
\includegraphics[width=1.0\textwidth]{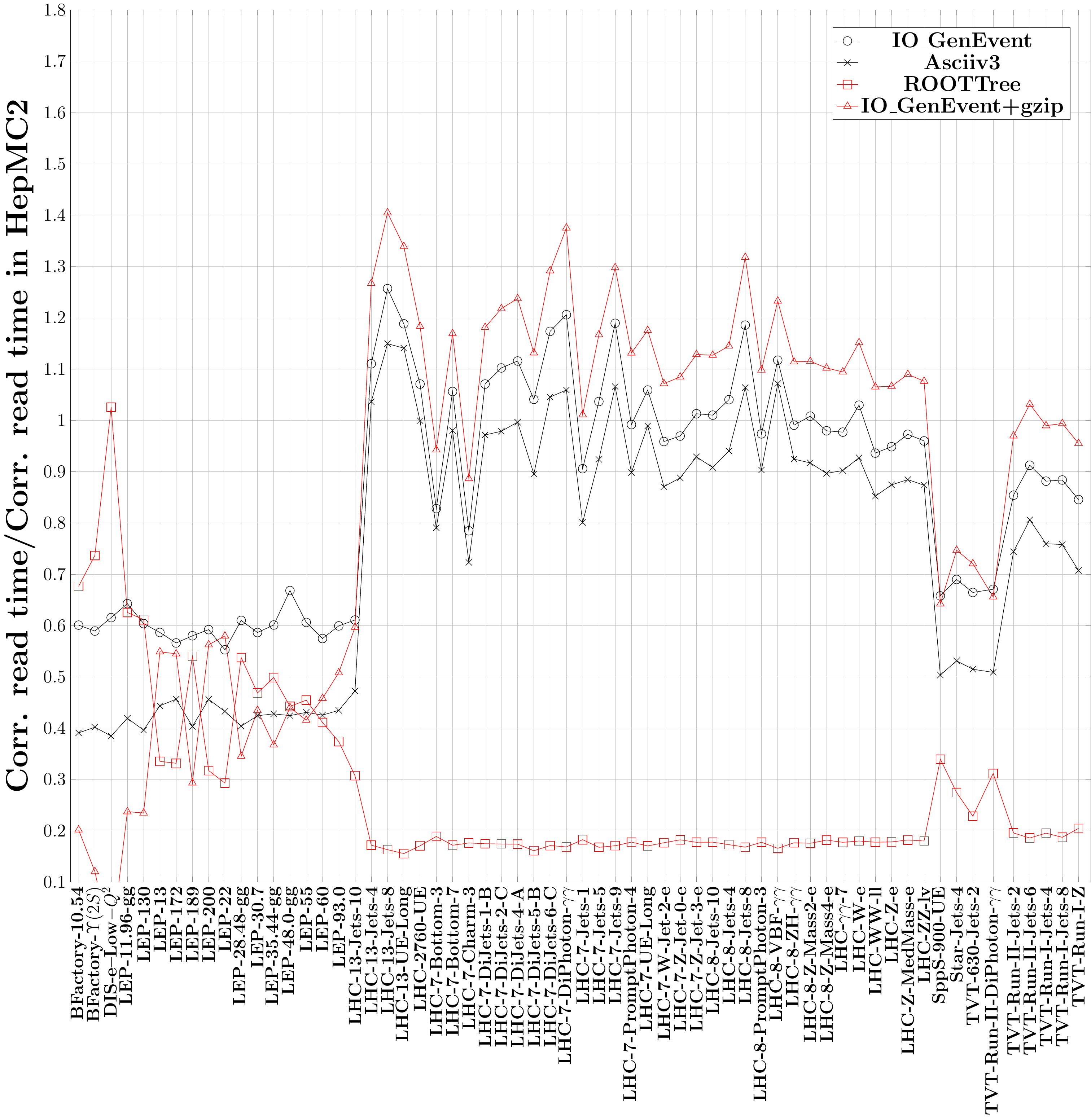}
\caption{Corrected reading time  of events samples in different formats relatively to the corrected reading time of same sample in HepMC2 IO\_GenEvent format.
The correction is done subtracting the time needed to read the first event in the file. See Fig.~\ref{fig:size} for details.}
\label{fig:totaltimecorr}
\end{figure}

The Fig.~\ref{fig:size} shows that |Asciiv3| with default precision has the same
size as |IO_GenEvent|, and the |ROOTTree| format provides the most efficient
packing of events ahead of compression with |zlib|.  The measurements of total
reading time for the samples are given in Fig.~\ref{fig:totaltime}. The same
measurements as described above were corrected for the time of opening of files
are given in Fig.~\ref{fig:totaltimecorr}.

The Fig.~\ref{fig:totaltimecorr} shows that reading from |Asciiv3| is typically
faster than from IO\_GenEvent in HepMC3.  The reading from |Asciiv3| is in
HepMC3 is sometimes slightly slower than reading from |IO_GenEvent| in HepMC2.
The small difference can be explained with extra time needed to assure thread
safety.

The |ROOTTree| format provides the most efficient reading of events for almost
all cases.

\section{Interfaces, examples and documentation}

\subsection{Interfaces}
The presented library contains some interfaces to the MCEGs, which do not
ship the interfaces to HepMC3, see Tab.~\ref{tab:summary}.
These interfaces can be used instantly in the production or tests to
generate the Monte Carlo simulated events.
One important difference between the HepMC2 and HepMC3 is that the later
delivers only it's interface for the Pythia6 generator, while the former
provided \cxx wrappers to the Pythia6 functions.

\subsection{Examples}
\label{examples}
For the users convenience, numerous example programs are provided with the library.
A brief overview of these codes is given in Tab.~\ref{tab:examples}.

\tabexamples

These examples can be modified and/or compiled using with external HepMC3 installation.
For instance, with an installed HepMC3 it is possible to compile examples only:
\begin{lstlisting}[language=bash]
mkdir -p myexamples
cd myexamples
git clone https://gitlab.cern.ch/hepmc/HepMC3
...
cd HepMC3/examples/
cmake -DUSE_INSTALLED_HEPMC3=ON CMakeLists.txt
cmake --build .
\end{lstlisting}

\vspace{1ex}

\subsection{Documentation}
\label{docs}
The online documentation is available on the HepMC3 home page~\cite{hepmc3}.  It
includes the automatically generated documentation on the codes as well as extra
material on specific topics, e.g.\ the LHEF format.  The same documentation can
be generated from the sources using the doxygen~\cite{doxygen} utility and
appropriate configuration options, e.g.\ %
\begin{lstlisting}[language=bash]
cmake -DHEPMC3_BUILD_DOCS=ON <other options> CMakeLists.txt
\end{lstlisting}

\vspace{1ex}

\subsection{Python bindings}
%\FloatBarrier

HepMC includes \cxx codes for Python~\cite{Rossum:1995:PT:869378} language
bindings. The codes are suitable for compilation of Python modules for
Python2.7~\cite{python27} and Python3~\cite{python3}.  These codes were
generated automatically using the binder~\cite{binder} utility and depend on the
pybind11~\cite{pybind11} header-only library included in the HepMC3 codes.  So
far the binding codes are available for all classes in HepMC3 and LHEF name
spaces but some in Search engine.  For usage examples please look into the
tests.  To turn on the compilation of bindings use
\begin{lstlisting}[language=bash]

cmake -DHEPMC3_ENABLE_PYTHON=ON <options> CMakeLists.txt
\end{lstlisting}
By default the Python modules will be generated for Python2 and Python3 if these
are found in the system.  The exact desired Python version can be specified
appropriate configuration options, e.g.\
\begin{lstlisting}[language=bash]

cmake -DHEPMC3_PYTHON_VERSIONS=2.7,3.4,3.6 <other options> CMakeLists.txt
\end{lstlisting}
In case the test suite is enabled, tests of python bindings with all the enabled
versions will run as well.  In the automatically generated codes it was assumed
that \lstinline{std::ostream} will be mapped onto \lstinline{io.stringIO()} and
similar objects.  The constructors of classes derived from
\lstinline{Reader/Writer} with \lstinline{std::ifstreams/std::ostreams} were
omitted.  To  benchmark the implemented capabilities, the Pythia8 HepMC3 interface
was re-implemented in Python and tested together with Python bindings of Pythia8, see Fig.\ref{fig:pypy}.

\begin{figure}[htbp]\centering
\includegraphics[width=1.0\textwidth]{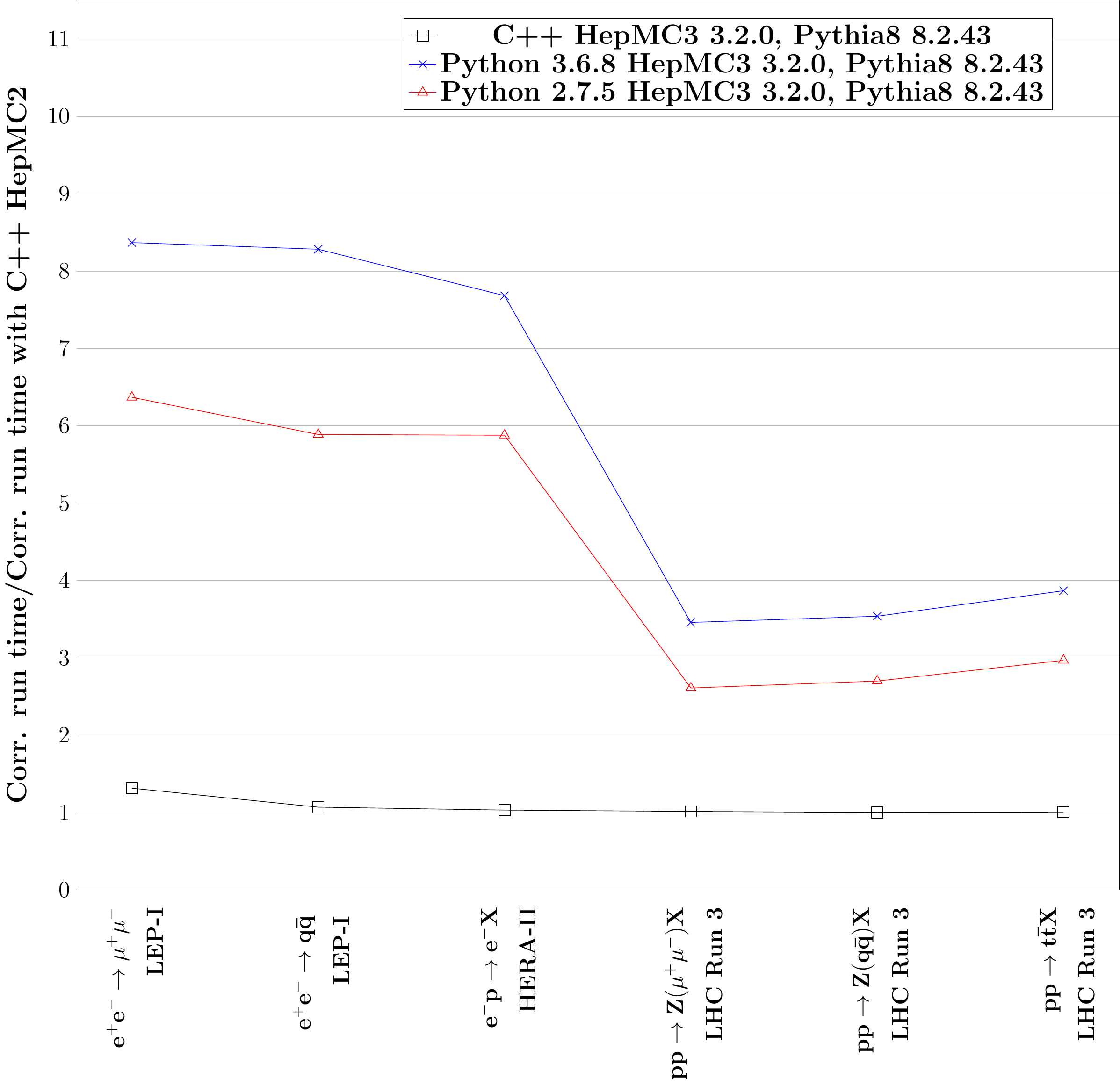}
\caption{Time needed to produce a sample of simulated events using different
Pythia8 interfaces to event record libraries relatively to the time needed
to produce a sample of same size using Pythia8 HepMC2 C++ interface.
The ratios were obtained from the measurements of time needed to produce
${\cal O}(10^4-10^5)$ events and were corrected for the effects of start-up in the same way as the
ratios in Fig.~\ref{fig:totaltimecorr}. The measurements were provided using the HepMC2, HepMC3, Pythia8 and python
packages from the EPEL repository on x86\_64 machine running CentOS7. The C++ codes were compiled with standard options
for this platform using the gcc compiler of version 4.8.5.
 }
\label{fig:pypi}
\end{figure}

Despite not being recommended, it should be possible to compile the Python
bindings using the installed version of HepMC3.  To do this, copy the |python|
directory outside the source tree, uncomment the line
\begin{lstlisting}[language=bash]

project(pyHepMC3 CXX)
\end{lstlisting}
in \myfile{python/CMakeLists.txt} and run CMake inside the |python| directory
with the option \lstinline{-DUSE_INSTALLED_HEPMC3=ON}.

The package {pyhepmc/pyhepmc-ng}~\cite{pyhepmcng} provides bindings to some core
functions of HepMC3.

\FloatBarrier

\section{Conclusions}

The HepMC3 library is designed to perform manipulations with event records of
High Energy Physics Monte Carlo Event Generators (MCEGs). The library version
3.2.0 has been released in November 2019.

The I/O functionality of the library has been extended to support common input
and output formats of HEP MCEGs, including formats used in Fortran HEP MCEGs,
formats used in HepMC2 library and ROOT.  The library is already supported by
many MCEGs (e.g.\ Herwig, Sherpa, WHIZARD), provides interfaces to others (Pythia8, TAUOLA and PHOTOS)
 and can replace the older HepMC versions in various applications dealing with Monte Carlo event records (e.g.\ in Rivet).
\section*{Acknowledgements}
We are grateful to our users that have helped us to find  and  fix problems in the library, improve the code and the documentation.

This work was supported in part by the European Union as part of the Marie Sklodowska-Curie Innovative Training Network MCnetITN3 (grant agreement no. 722104).
AB thanks The Royal Society for University Research Fellowship grant UF160548, and the University of Glasgow for funding through the Leadership Fellow scheme.
LL was supported in part by the Swedish Research Council, contract number 2016-03291.
\newpage

\begin{appendices}
\section{}
\subsection{Status codes}
\FloatBarrier
\label{status}
\tabstatusp

\tabstatusv

\FloatBarrier
\subsection{Compatibility with earlier version of HepMC3}
\label{compatibility}
Prior to version 3.2.0, the following attributes were handled during the reading
of |IO_GenEvent| files in a different way.
The differences are:\\
For the particles:
\begin{itemize}
\item The particle flows were added to the event  as multiple \lstinline{IntAttribute} attributes with names ``flow1'' and  ``flow2''.
\item The vertex weights were added to the event  as  multiple \lstinline{DoubleAttribute} attributes with names  ``weights1'', ``weights2'' \dots ``weightsN''.
\item The event random number generator states were added to the event  as multiple \lstinline{IntAttribute} attributes with names  ``random\_state1'', ``random\_state2'' \dots ``random\_stateN''.
\end{itemize}

The old behaviour during the event reading can be restored setting the options
\lstinline{"particle_flows_are_separated"},
\lstinline{"vertex_weights_are_separated"} and\\
\lstinline{"event_random_states_are_separated"}.

\subsection{Attributes}
\label{attributes}
The attributes described below have a simple structure with all important members being public.
Therefore, the functions like \lstinline{void GenPDFInfo::set(...)} are provided only for convenience and are not described in detail below.

\subsubsection{GenPdfInfo}
The \lstinline{GenPDFInfo} contains the following data members:
\begin{itemize}
\item \lstinline{int parton_id[2]} -- array with two elements holding PDG I.D.  for the first and second interacting parton.
\item \lstinline{int pdf_id[2]} -- array with two elements  holding I.D.s of PDF distributions  as encoded in the LHAPDF~\cite{Buckley:2014ana} library.
\item \lstinline{double scale} -- value of factorisation scale (in \GeV).
\item \lstinline{double x[2]} --  array with two elements holding fractions of interacting partons momentum  with respect to the momentum of their beams.
\item \lstinline{double xf[2]} -- array with two elements holding  the values of PDF.
\end{itemize}
The representation of \lstinline{GenPDFInfo} as \lstinline{std::string} is structured as\\[1ex]
\mybox{parton\_id[0]} \mybox{parton\_id[1]} \mybox{x[0]} \mybox{x[1]} \mybox{scale} \mybox{xf[0]}  \mybox{xf[1]}  \mybox{pdf\_id[0]}   \mybox{pdf\_id[1]}

\subsubsection{GenCrossSection}
The \lstinline{GenCrossSection} contains the following data members:
\begin{itemize}
\item \lstinline{long int accepted_events}       -- the number of generated events.
\item \lstinline{long int attempted_events}      -- the number of  attempted  events.
\item \lstinline{std::vector<double> cross_sections}    -- values of cross-sections.
\item \lstinline{std::vector<double> cross_section_errors} -- values of  cross-section uncertainties.
\end{itemize}
The representation of \lstinline{GenCrossSection} as \lstinline{std::string} is structured as\\[1ex]
\mybox{cross\_sections[0]} \mybox{cross\_section\_errors[0]} \mybox{accepted\_events} \mybox{attempted\_events}\\
\mybox{cross\_sections[1]} \mybox{cross\_section\_errors[1]} \dots

\subsubsection{GenHeavyIon}
The \lstinline{GenHeavyIon} contains the following data members:
\begin{itemize}
\item \lstinline{int Ncoll_hard}  the number of hard nucleon-nucleon collisions.
\item \lstinline{int Npart_proj}  the number of participating nucleons in the projectile.
\item \lstinline{int Npart_targ} the number of participating nucleons in the target.
\item \lstinline{int Ncoll}  the number of inelastic nucleon-nucleon collisions.
\item Deprecated \lstinline{int spectator_neutrons}  Total number of spectator neutrons.
\item Deprecated  \lstinline{int spectator_protons} Total number of spectator protons.
\item \lstinline{int N_Nwounded_collisions} Collisions with a diffractively excited target nucleon.
\item \lstinline{int Nwounded_N_collisions}  Collisions with a diffractively excited projectile nucleon.
\item \lstinline{int Nwounded_Nwounded_collisions}  Non-diffractive or doubly diffractive collisions.
\item \lstinline{double impact_parameter}  The impact parameter.
\item \lstinline{double event_plane_angle}  The event plane angle.
\item Deprecated \lstinline{double eccentricity}  The eccentricity.
\item \lstinline{double sigma_inel_NN}  The assumed inelastic nucleon-nucleon cross section.
\item \lstinline{double centrality}  The centrality.
\item \lstinline{double user_cent_estimate}  A user defined centrality estimator.
\item \lstinline{int Nspec_proj_n}  The number of spectator neutrons in the projectile.
\item \lstinline{int Nspec_targ_n}  The number of spectator neutrons in the target.
\item \lstinline{int Nspec_proj_p}  The number of spectator protons in the projectile.
\item \lstinline{int Nspec_targ_p} The number of spectator protons in the target.
\item \lstinline{std::map<int,double> participant_plane_angles} Participant plane angles.
\item \lstinline{std::map<int,double> eccentricities} Eccentricities.
\end{itemize}
The \lstinline{std::string} representation of \lstinline{GenHeavyIon}
can be built in two ways:
\begin{itemize}
\item
The ``old'' version is structured as:\\[1ex]
\mybox{v0} \mybox{Ncoll\_hard}  \dots  \mybox{Nspec\_targ\_p}\\
\mybox{participant\_plane\_angles.size()}\\
\mybox{participant\_plane\_angles[0].first} \mybox{participant\_plane\_angles[0].second} \dots\\
\mybox{eccentricities.size()}\\
\mybox{eccentricities[0].first}   \mybox{eccentricities[0].second} \dots\\[1ex]
With all other members described above listed between \mybox{Ncoll\_hard}  and  \mybox{Nspec\_targ\_p}.
\item
The ``new'' version is structured as:\\[1ex]
\mybox{v1} \mybox{Ncoll\_hard}  \dots  \mybox{Nspec\_targ\_p}\\
\mybox{participant\_plane\_angles.size()}\\
\mybox{participant\_plane\_angles[0].first} \mybox{participant\_plane\_angles[0].second} \dots\\
\mybox{eccentricities.size()}\\ \mybox{eccentricities[0].first} \mybox{eccentricities[0].second} \dots\\[1ex]
With all non-deprecated members described above listed between \mybox{Ncoll\_hard}  and  \mybox{Nspec\_targ\_p}.
\end{itemize}

The ``new'' should comply to the Lisbon Accord~\cite{LA} and is aimed to be
helpful for the groups performing heavy-ion physics studies.

\end{appendices}

\FloatBarrier
%\newpage

\section*{References}
%{\bibliographystyle{./HepMC3}{\raggedright\bibliography{HM3.bib}}}\vfill\eject  %Bibliography
{\bibliographystyle{elsarticle-num}\bibliography{HM3.bib}}\vfill\eject  %Bibliography
\end{document}